\pdfoutput=1
\documentclass[10pt,journal]{IEEEtran}

\usepackage{graphicx,amssymb}
\graphicspath{{figures/}}
\usepackage{float}
\usepackage{wrapfig}
\usepackage{fancyhdr}    
\usepackage{fancybox}
\usepackage{amsmath}
\usepackage[hyphens]{url}
\usepackage{latexsym,amsmath,caption,epsfig,cite,array}
\usepackage{circuitikz}
\usepackage{mathrsfs}   
\usepackage{verbatim}
\usepackage{pgfplots} 
\usetikzlibrary{patterns,pgfplots.fillbetween}
\pgfplotsset{compat=1.3, tick style={black}}
\usepackage{multirow}
\usepackage{xcolor}
\usepackage[font=footnotesize,caption=false,subrefformat=parens,labelformat=parens]{subfig}
\usepackage{algorithm}
\usepackage{algorithmic}
\usepackage[english]{babel}
\usepackage[utf8]{inputenc}
\usepackage{cutwin,lipsum}
\usepackage{soul}
\usepackage{nomencl}
\makenomenclature

\usepackage{hyperref}
\hypersetup{
    colorlinks=true,
    linkcolor=blue,     
    urlcolor=cyan,
    }
\definecolor{LightCyan}{rgb}{0.88,1,1}

\tikzset{
every pin/.style={fill=yellow!50!white,rectangle,rounded corners=3pt,font=\footnotesize},
small dot/.style={fill=yellow!50!red,circle,scale=0.75}
}

\title{
Controller Hardware-in-the-loop (C-HIL) Testing of Decentralized EV-Grid Integration
}
\author{\IEEEauthorblockN{Calvin Flack, Emin Ucer, Charles Parker Smith, Mithat Kisacikoglu}
\\
\IEEEauthorblockA{Dept. of Electrical and Computer Engineering,
The University of Alabama, Tuscaloosa, AL}\\
E-mails: \{eucer, caflack, cpsmith4\}@crimson.ua.edu, mkisacik@ua.edu

\thanks{This material is supported by the National Science Foundation under Award No~1755996.}
}

\begin{document}

\IEEEoverridecommandlockouts

\IEEEpubid{\begin{minipage}[t]{\textwidth}\ \\[10pt]
        \centering\footnotesize
        {~\copyright2022 IEEE Personal use of this material is permitted.  Permission from IEEE must be obtained for all other uses, in any current or future media, including reprinting/republishing this material for advertising or promotional purposes, creating new collective works, for resale or redistribution to servers or lists, or reuse of any copyrighted component of this work in other works.}
\end{minipage}}

\maketitle

\vspace{-10mm}
\begin{abstract}
\begin{small}
Present grid infrastructure is unprepared for the wide-scale integration of electric vehicles (EVs). Real-time grid simulation and hardware testing are necessary for the fast and accurate development of EV charging control strategies that are applicable in the field to mitigate their adverse effects. To help facilitate this goal, in this study, we develop a controller verification testbed using a baseline bidirectional AC/DC converter, commonly used in on-board EV chargers, within a hardware-in-the-loop (HIL) test system. The converter runs real-time on a very-fast time scale and integrates into an external grid simulator. We utilize the strength of the HIL device in simulating power electronics while simultaneously realizing a complex distribution grid operation in real-time. The whole system holistically provides an opportunity to see the implementation benefits of a decentralized charging algorithm on the voltage stability of the distribution grid. The study presents a step-by-step approach to developing and testing controllers, providing an essential contribution in the field demonstration of decentralized EV charger control strategies.
\end{small}
\end{abstract}
\vspace{-4mm}
\section{Introduction}


Electric vehicles (EVs) will become more common in the coming years in the constant search for cleaner transportation. While EVs provide an alternative to traditional internal combustion engine (ICE) vehicles, they can significantly strain the distribution grid operation if mass integrated. An on-board EV charger can operate at a rate as high as the peak power consumption of a typical house~\cite{HomeEnergyUsage,HomeEnergyModeling,ChargerPowerUsage}. Today, more than 80\% of EV owners prefer to charge their EVs at home in the US~\cite{EVowners}. In a world where most homes have an EV, it would easily overload the present grid equipment if the power demand were to nearly double as people return to their homes and start charging their vehicles. Smart and coordinated charging methods will be critical for the smooth integration of EVs on a mass scale. 

While EVs can disturb the grid, they can also benefit it if adequately integrated. With minimal hardware modifications~\cite{kisacikoglu2013bidirectional,kisacikoglu2015single}, EV chargers can provide grid reactive power services to compensate for nearby inductive loads. Furthermore, studies show how EVs can filter the distortion created from other nearby loads~\cite{Taghizadeh2018THDCompensation}. Other studies explore altering the charging power to compensate for voltage drop during times of peak power usage to provide decentralized demand response to the grid~\cite{ucer2019analysisPESGM}. 


Unidirectional EV chargers can only compensate for grid voltage drops by lowering their charging power. No more compensation can be provided if the drop is severe enough that the charging power reaches a minimum. Bidirectional chargers allow full compensation as they are capable of supplying real power as well. Additionally, four-quadrant bidirectional chargers add the ability to inject or absorb reactive power in addition to real power. With the high power capability of EV chargers, bidirectional chargers could easily supply enough power to support short-term increased power demand. Whether it is a unidirectional or bidirectional, for all the ancillary services provided to the grid, the developed control algorithms should be practical and easily implemented in the field. 

Hardware-in-the-loop (HIL) devices are used to simulate the function of a system in real-time. This way, power/control hardware external to the system can be tested. This is done by providing feedback from the simulated system to the external hardware while receiving signals from the hardware to influence the simulated system. HIL devices increase controller development speed and remove the need to build the full system or risk damage to components due to controller issues. These advantages have made HIL devices quite popular for their simplicity and ease of use. Past work using HIL devices showcased grid-tied AC/DC converter controllers and their impact on basic grid models~\cite{Bagudai2019Evaluation,Chattopadhyay2014Comparison}.

A decentralized Additive Increase Multiplicative Decrease (AIMD) charging algorithm was selected to increase grid stability because of its simplicity and ability to scale~\cite{ucer2020decentralized}. The literature on decentralized charging algorithms mainly focuses on simulated results \cite{Liu2015}. These studies provide good insight into new strategies, but simulation alone can potentially omit real-world problems associated with controller implementation. There has also been limited work in real-time control algorithm validation, but with control algorithms different than AIMD \cite{Marinescu2018}. Additionally, most proposed work utilizes simple grid models \cite{Lakshminarayanan2019,Jayawardana2021} whereas we present a complex distribution system, modeling both primary and secondary sides down to a total of $320$ end-nodes that have real power consumption profiles. This presents the opportunity to build a system for the purpose of validating a purely decentralized AIMD charging algorithm in real-time.

This paper will investigate the integration of a bidirectional AC/DC converter model realized on a fast time-scale Typhoon HIL 402~\cite{TyphoonHIL402} simulator. A more complex grid simulation runs a versatile multi-core real-time digital simulator, OPAL-RT OP5600~\cite{OPALRT}. The controller will utilize a decentralized AIMD algorithm to mitigate adverse grid impacts. The overall implementation accuracy of the algorithm will be assessed using a TI C2000 micro-controller. 


\vspace{-2mm}
\section{System Description and Modeling}
\subsection{On-board Charger Power Electronics Topology}

\begin{figure}[tb]
\centering
\includegraphics[width=87mm]{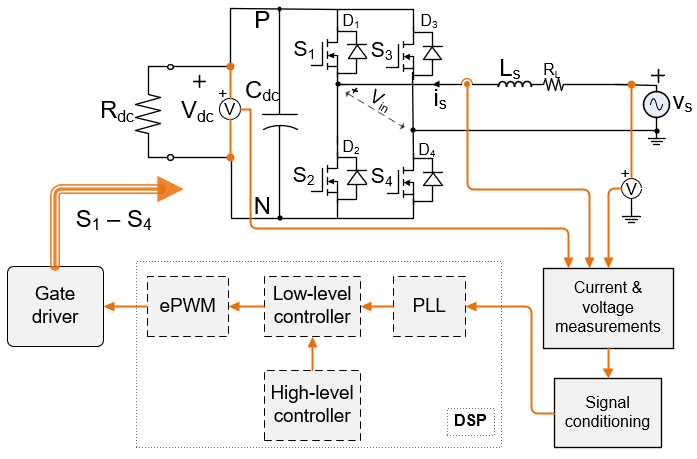}
\vspace{-2mm}
\caption{AC/DC converter topology schematic.} 
\label{fig:overallSchematic}
\vspace{-2mm}
\end{figure}

In typical two-stage on-board chargers, the AC/DC converter controls real and reactive power received from the grid, while the DC/DC stage manages the current sent to the EV battery. Since our focus is the grid-integration and its implementation using HIL, only the AC/DC stage is analyzed. Fig.~\ref{fig:overallSchematic} shows the baseline AC/DC converter topology and its control-loop developed for this study. It uses a full-bridge active front-end rectifier and a resistor connected to the DC output. This configuration can easily emulate the behaviour of an on-board charger for advanced grid services and simultaneously realize battery charging without needing an actual battery pack model. This is especially advantageous for future hardware testing of EV-grid integration without the need to maintain a fail-proof battery testing environment. The system design specifications are listed in Table~\ref{tab:system_param}. The parameters are selected to be compatible with a future actual hardware demonstration using wide band-gap SiC MOSFETs. 

\begin{table}[tb]
\caption{System Specifications}
\vspace{-1mm}
\label{tab:system_param}
 \centering
{
\begin{tabular}{l|cc}
\hline
Parameter & Symbol & Value \\

\hline 
\hline 
Charger apparent power&S &10 kVA\\
Grid voltage &$V_s$ &240 V\\
Grid frequency &$\omega_{0}$ &$2\pi60$  rad/s\\
Switching frequency &$f_{sw}$ &72 kHz\\
Filter Inductance &$L_{s}$ &500 $\mu$H\\
Output voltage &$V_{dc}$ &340-800 V\\
Output capacitance &$C_{dc}$ &500 $\mu$F\\
Output equivalent resistance&$R_{dc}$&60~$\Omega$\\

\hline
\end{tabular}
}
\vspace{-3mm}
\end{table}

\subsection{Modeling of the Converter Transfer Function}
\begin{figure}[h]
\vspace{-3mm}
\centering
\includegraphics[width=70mm]{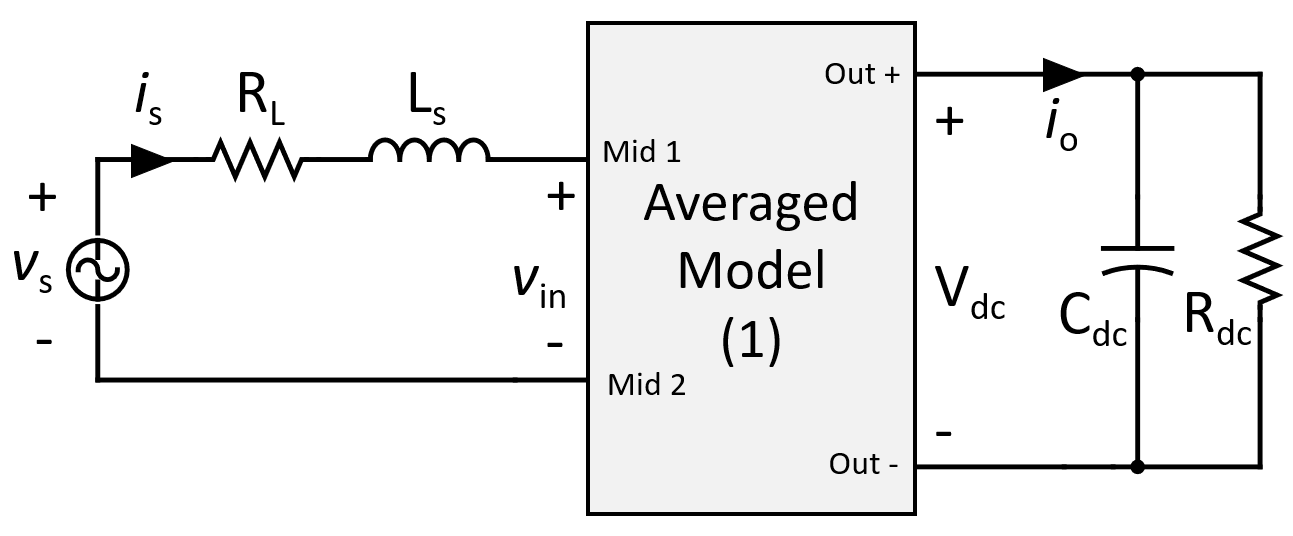}
\vspace{-2mm}
\caption{Average model of the AC/DC converter.} 
\label{fig:blackBoxSchematic}
\vspace{-6mm}
\end{figure}

The modeling and controller design approach follows a similar methodology presented in \cite{kisacikoglu2015single,Kisacikoglu2013dissertation}. This design employs a bipolar modulation technique, where the switches operate in pairs. When switches $S_1$ and $S_4$ are on, switches $S_2$ and $S_3$ are off. This means that the voltage $V_{in}$ is either $+V_{dc}$ or $-V_{dc}$, with no instance of $V_{in}=0$. If we replace the full bridge in Fig.\ref{fig:overallSchematic} with a black box that has AC and DC terminals, we get the average (free of switching devices) representation of the converter as depicted in Fig.\ref{fig:blackBoxSchematic}. To derive the relationship between the input current and the control input, the relationship between the AC and DC terminals must be defined.

The relationship between the
input and output currents and voltages can be found if the circuit is averaged over a switching period. The output capacitance ($C_{dc}$) is assumed to be high enough to keep the DC voltage ($V_{dc}$) constant. Assuming the full-bridge circuit is lossless, we can write:
%
\begin{equation} \label{eq:modulation}
\begin{split}
    v_{in} &= m\cdot V_{dc} \\
    i_{s} &= \frac{I_{o}}{m}
\end{split}
\end{equation}
where
\begin{equation} \label{eq:modulation}
\begin{split}
    m &= M\cdot sin(\omega_{0}t-\delta),\,\,\,   |M| \leq 1,\\
    v_s&=\sqrt{2}\,V_s\,sin(\omega_{0}t).
\end{split}
\end{equation}
$m$ is the sinusoidal modulation index, $M$ is the modulation index amplitude, $I_{o}$ is the output DC current flowing into the RC network, $i_s$ is the grid current,  and $w_{0}$ is the grid angular frequency. By applying KVL around the loop on the AC side of the circuit, the following differential equation is obtained:

\begin{equation} \label{eq:KLV}
    \frac{di_{s}}{dt} = -\frac{R_{L}}{L_{s}}i_{s} + \frac{1}{L_{s}}v_{s}-\frac{1}{L_{s}} \underbrace{v_{in}}_\text{m$V_{dc}$}.
\end{equation}

%
After taking the Laplace transform of both sides, the following expression is obtained:
\begin{equation} \label{eq:TF}
\begin{split}
sI_{s}(s)+\frac{R_{L}}{L_{s}}I_{s}(s) &= \frac{1}{L_{s}}V_{s}(s)-\frac{V_{dc}}{L_{s}}M(s) \\
I_{s}(s)\left(s+\frac{R_{L}}{L_{s}} \right) &= \frac{1}{L_{s}}V_{s}(s)-\frac{V_{dc}}{L_{s}}M(s) \\
I_{s}(s) &= \frac{1}{R_{L}+sL_{s}}V_{s}(s)-\frac{V_{dc}}{R_{L}+sL_{s}}M(s).
\end{split}
\end{equation}
Here, the transfer function between the input current $i_{s}$ and the control input $m$ is found to be
\begin{equation}\label{eq:avg_model}
    G_P(s)=\frac{I_{s}(s)}{M(s)} = -\frac{V_{dc}}{R_{L}+sL_{s}}.
\end{equation}
%


We should note that $v_{s}$ is a sinusoidal voltage, and it is fully coupled with the system. Therefore, we can model it as a sinusoidal disturbance to the system that is supposed to be eliminated by the controller.

\section{Controller Design}


This design consists of two controller stages. The first stage is the inner-loop current controller, and the second stage is comprised of the outer-loop active power (P), reactive power (Q), and DC bus voltage ($V_{dc}$) controllers. The controller meets the P demand via dynamically adjusting the dc-link voltage between 340-800~V.  

\begin{figure}[tb]
\centering
\includegraphics[width=80mm]{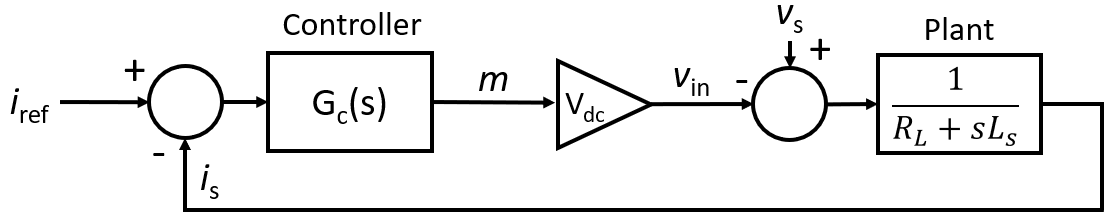}
\vspace{-2mm}
\caption{Closed-loop block diagram of the system.\footnotesize}
\label{fig:inner_controller}
\vspace{-5mm}
\end{figure}

The current controller could be designed by considering the system’s average model \eqref{eq:avg_model} transfer function, and the outer loop parameters can be tuned via empirical trial and error methods. 
The block diagram of the designed system is shown in Fig.~\ref{fig:inner_controller}. Plant refers to the converter transfer function and $G_{c}(s)$ is the current controller. The reference input ($i_{ref}$) to the system will be a sinusoidal signal. The disturbance ($v_{s}$) is also sinusoidal at the same frequency as the reference. Therefore, the controller ($G_{c}(s)$) must have a very large gain at $\omega_{0}$ to reject all the disturbances and eliminate the steady-state errors.

One controller structure that is known to have a large gain at a specific frequency is the proportional resonant (PR) controller. The transfer function of a typical PR controller is shown below.

\vspace{-1mm}
\begin{equation}
    G_{c}(s) = K_{p} + \frac{2K_{i}\omega_{c}}{s^{2}+2\omega_{c}s+\omega_{0}^{2}}.
\end{equation}

\begin{figure}[b]
\vspace{-6mm}
\centering
\includegraphics[width=80mm]{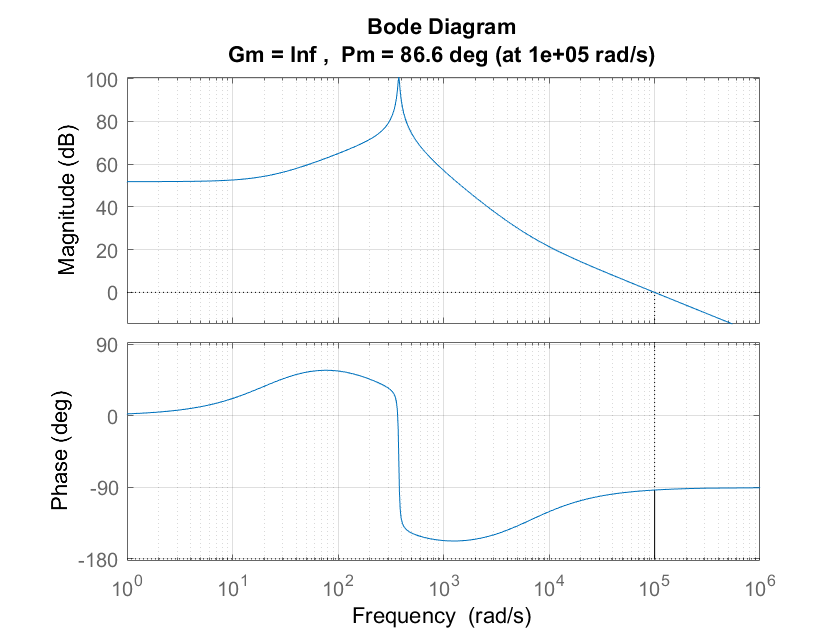}
\vspace{-2mm}
\caption{Final loop transfer function frequency response} 
\label{fig:modified_PR}
\end{figure}

To tune the controller, first an arbitrary and reasonable pair of $K_{p}$ and $K_{i}$ are chosen. In this design, we set $K_{p}=1$ and $K_{i}=500$, and $\omega_{c}$ is chosen to be $2\pi$ rad/s. Although these coefficients result in a stable response with infinite gain margin and positive phase margin, the bandwidth (cross-over frequency) of the system is too high for the digital controller to respond (i.e., $1$~Mrad/s). 
After inserting a gain of 0.1 in the controller path to slow down the response, the cross-over frequency of the new system with the PR controller is now $100$~krad/s. The new ($G_{c}$) then becomes:

\vspace{-1mm}
\begin{equation}
    G_{c}(s) = \frac{-0.1s^2-629.6s-1.421\cdot10^{4}}{s^2+12.57s+1.421\cdot10^{5}}.
\end{equation}
The resulting frequency response of the loop transfer function $G_L(s)=G_C(s)\times G_P(s)$ is shown in Fig.~\ref{fig:modified_PR}. 

This continuous domain transfer function is converted to discrete-time using a sampling frequency of 72~kHz: \eqref{eq:discrete_controller}.
\begin{equation} \label{eq:discrete_controller}
    G_{c}(z) = \frac{-0.1z^2+0.1913z-0.09126}{z^2-2z+0.9998}.
\end{equation}

The outer loop that handles the active and reactive power tracking should be slower than the inner current controller. The P-Q theory is a commonly used method to implement power measurements in a discrete environment~\cite{PQtheory}. With this method, the voltage and current measurements ($V_{\alpha}$ and $I_{\alpha}$) are first delayed a quarter of a grid cycle to obtain the quadrature components ($V_{\beta}$ and $I_{\beta}$). The P and Q values are then computed using a non-linear combination of the delayed and present voltage and current components:
\begin{equation} \label{eq:PQ}
\begin{split}
    P &= 0.5\times(V_{\alpha}I_{\alpha}+V_{\beta}I_{\beta}) \\
    Q &= 0.5\times(V_{\alpha}I_{\beta}-V_{\beta}I_{\alpha})
\end{split}
\end{equation}
To eliminate the ripples on the power measurements, a discrete low pass filter (LPF) with a cut-off frequency of 20~Hz is used after the calculation of the quadrature components. For a given P and Q reference, the resulting input current is calculated as:
\begin{equation} \label{eq:current}
    i_{ref} = \sqrt{2}I_{s}sin(\omega_{0}t+\theta)
\end{equation}
where
\vspace{-4mm}
\begin{equation}
    \begin{split}
        I_{s}&=\frac{P_{ref}}{V_{s}cos(\theta)} \\
        \theta &= tan^{-1} ( \frac{Q_{ref}}{P_{ref}})
    \end{split}
\end{equation}

\begin{figure}[bt]
\centering
\includegraphics[width=85mm]{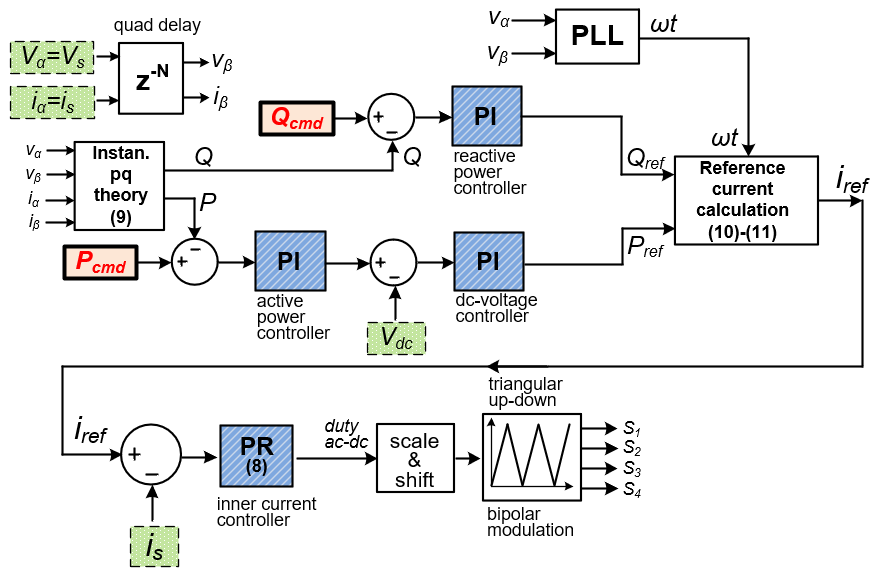}
\vspace{-2mm}
\caption{System controller diagram for inner and outer loops~\cite{kisacikoglu2015single}.} 
\label{fig:controller}
\vspace{-5mm}
\end{figure}

The outer loops are controlled by discrete PI controllers. The coefficients of the PI controllers for V-loop, P-loop and Q-loop shown in Fig.~\ref{fig:controller} are: $K_p^V$=0.1, $K_i^V$=20, $K_p^P$=2.5, $K_i^P$=2.5, $K_p^Q$=0.1, and $K_i^Q$=20.

\section{Real-time HIL Simulation Analysis}

\subsection{Dynamic Analysis of the Converter Operation}
\begin{figure}[bt]
\vspace{-3mm}
\centering
\includegraphics[width=83mm]{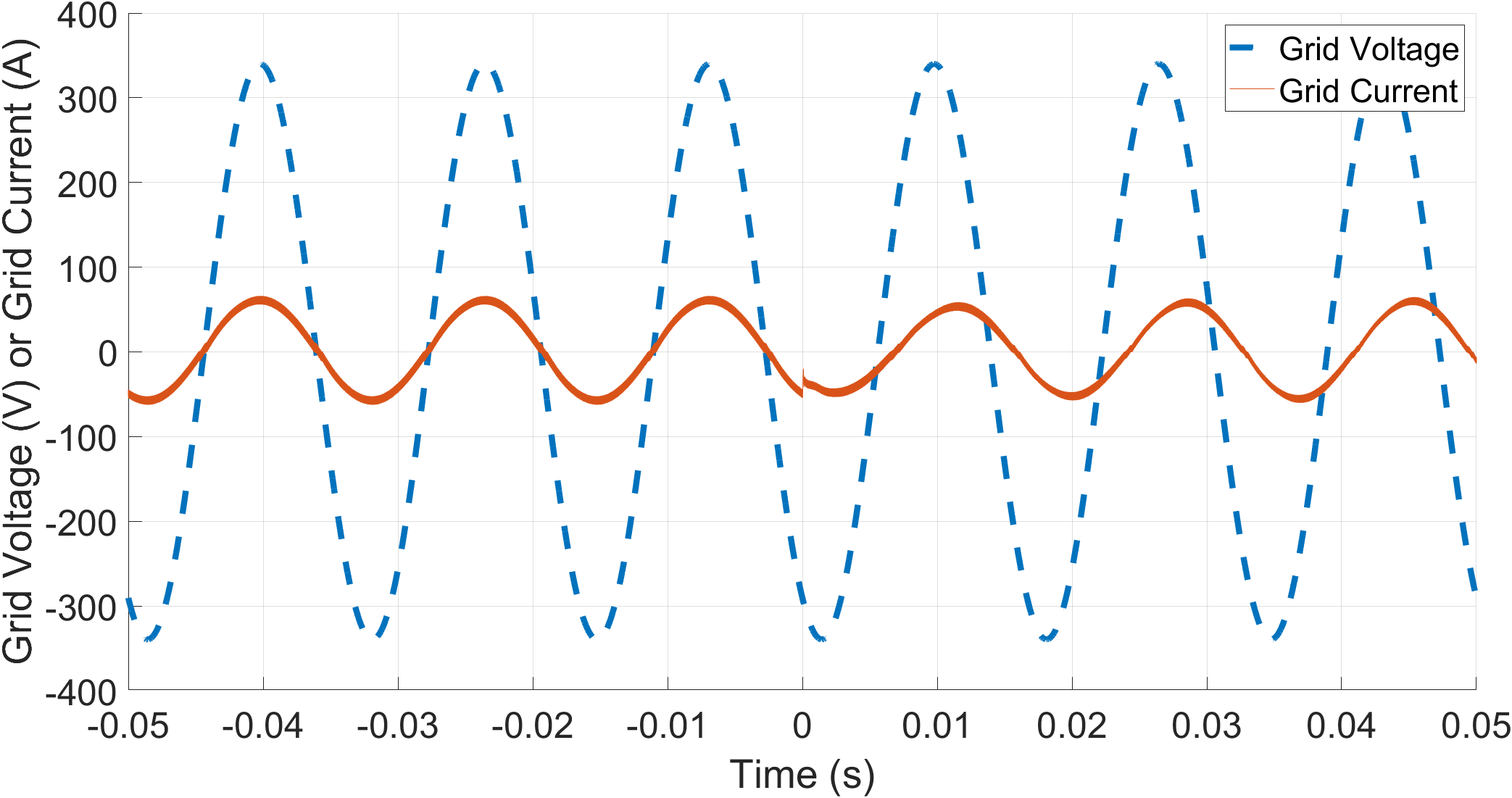}
\vspace{-2mm}
\caption{Transient test 1: grid voltage and current.}
\label{fig:TransientTest1Current}
\vspace{-1mm}
\end{figure}

\begin{figure}[bt]
\vspace{-3mm}
\centering
\includegraphics[width=83mm]{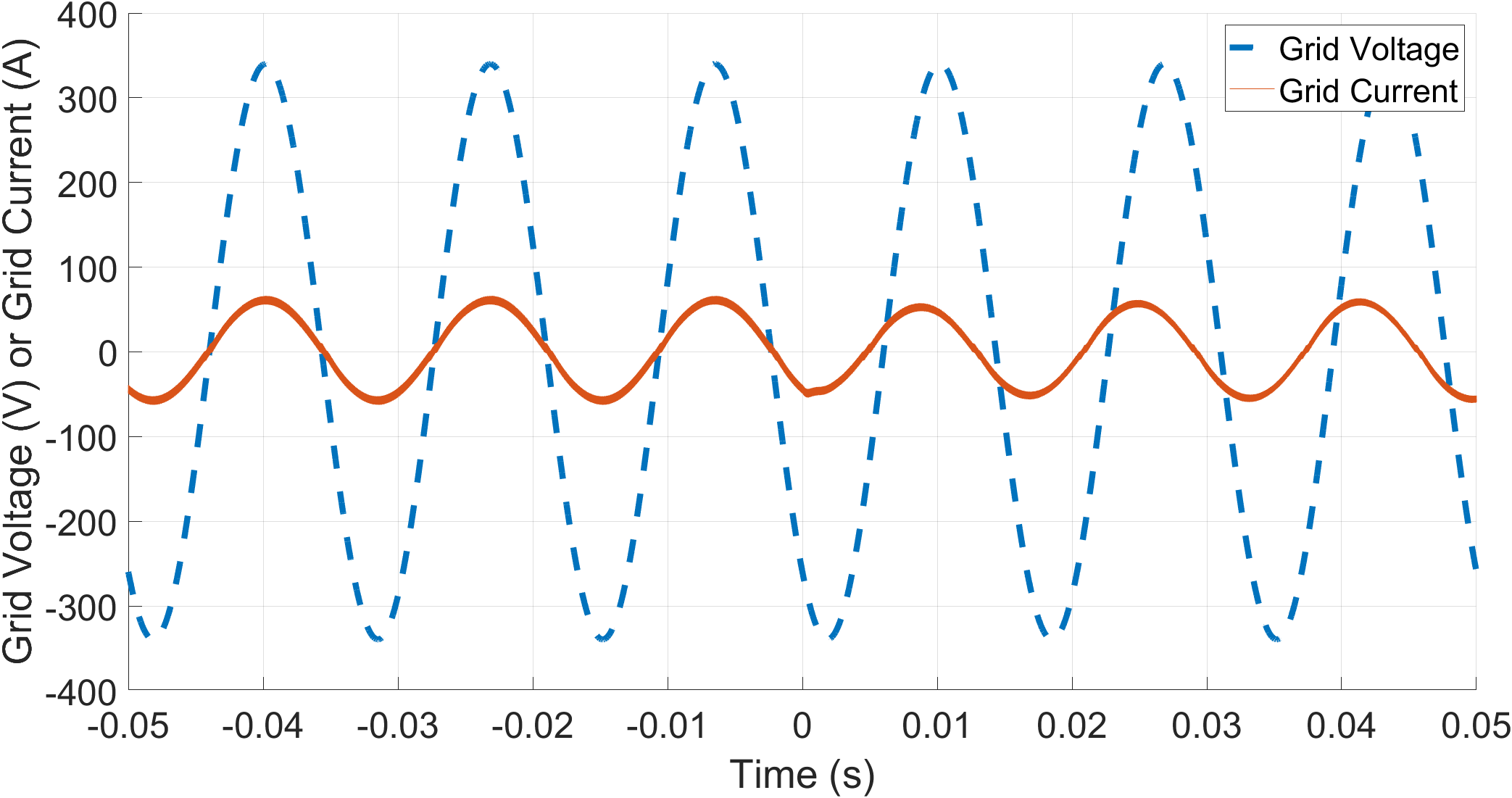}
\vspace{-2mm}
\caption{Transient test 2: grid voltage and current.}
\label{fig:TransientTest2Current}
\vspace{-5mm}
\end{figure}

The controller was implemented in a TI TMS320F28335 DIMM100 based control card and tested using the Typhoon HIL 402 environment. We used an interface card to connect the DSP to Typhoon HIL 402. There are two test scenarios implemented to analyze the transient performance of the charger. Each test began with the converter operating at rated apparent power and unity power factor ($P_{ref}{=}10$~kW and $Q_{ref}{=}0$~kVAR). Then, a step response was provided at $t{=}0~s$ to shift the desired converter output to two different power factors still at rated power, one is 0.707 pf lagging, and the other is 0.707 pf leading. Namely, Test\#1 has $P_{ref}{=}10{\rightarrow}7.07$~{kW} and  $Q_{ref}{=}0{\rightarrow}7.07$~kVAR; and Test\#2 has $P_{ref}{=}10{\rightarrow}7.07$~kW and $Q_{ref}{=}0{\rightarrow}-7.07$~kVAR. Resulting grid voltage and current can be seen in Figs.~\ref{fig:TransientTest1Current}-\ref{fig:TransientTest2Current} for the two tests, showing successful transient performance. 




\subsection{Grid Integration Simulation}
\begin{figure}[bt]
\vspace{-6mm}
\centering
\includegraphics[width=90mm]{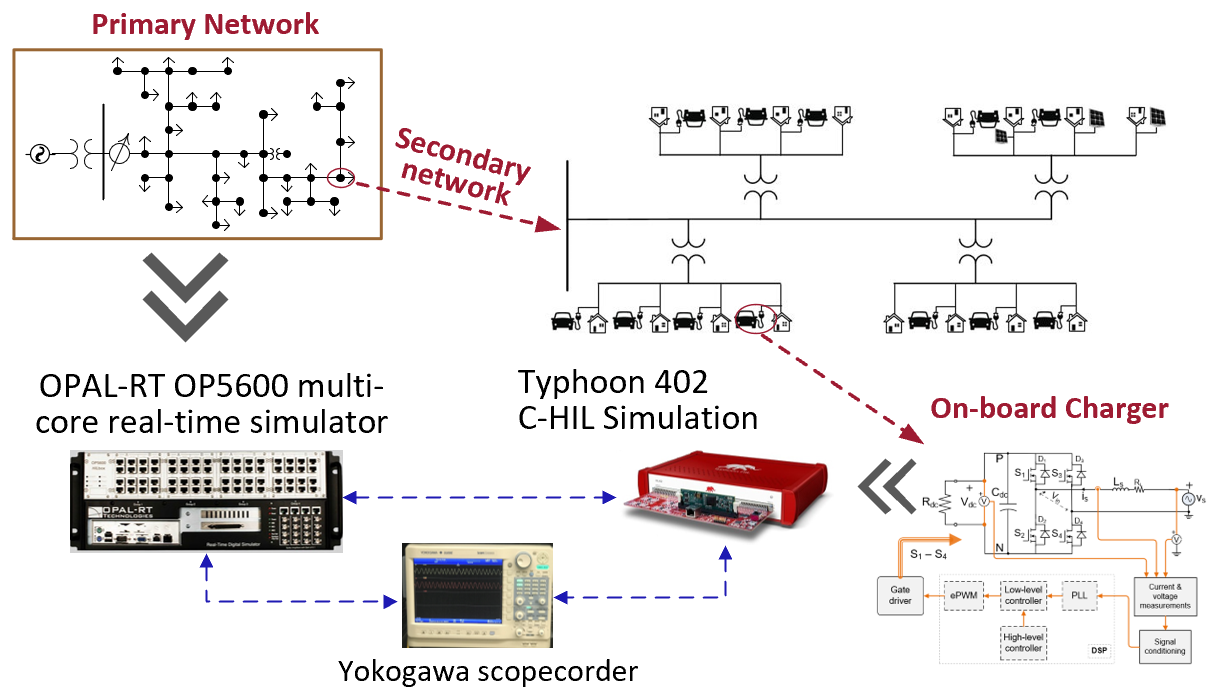}
\vspace{-4mm}
\caption{Overview of EV grid integration test system.}
\label{fig:gridSimDiagram}
\end{figure}

The on-board charger, implemented in Typhoon HIL 402, is integrated into a real-time grid simulator using OPAL-RT OP5600. The real-time grid simulation environment and its implementation are shown in Fig.~\ref{fig:gridSimDiagram}. It features a 2.5~MVA, 230~kV/4.8~kV, 37-bus, three-phase balanced network for the distribution grid that is composed of residential house and EV loads. In total, this grid model has 10 neighborhoods with 16 homes in each neighborhood. The EV charger is operating in one of the 16 homes within a neighborhood. There are a total of $320$ end-nodes. A diagram of the total HIL test system is shown in Fig.~\ref{fig:gridSimDiagram}.

\begin{figure}[bt]
\vspace{-3mm}
\centering
\includegraphics[width=68mm]{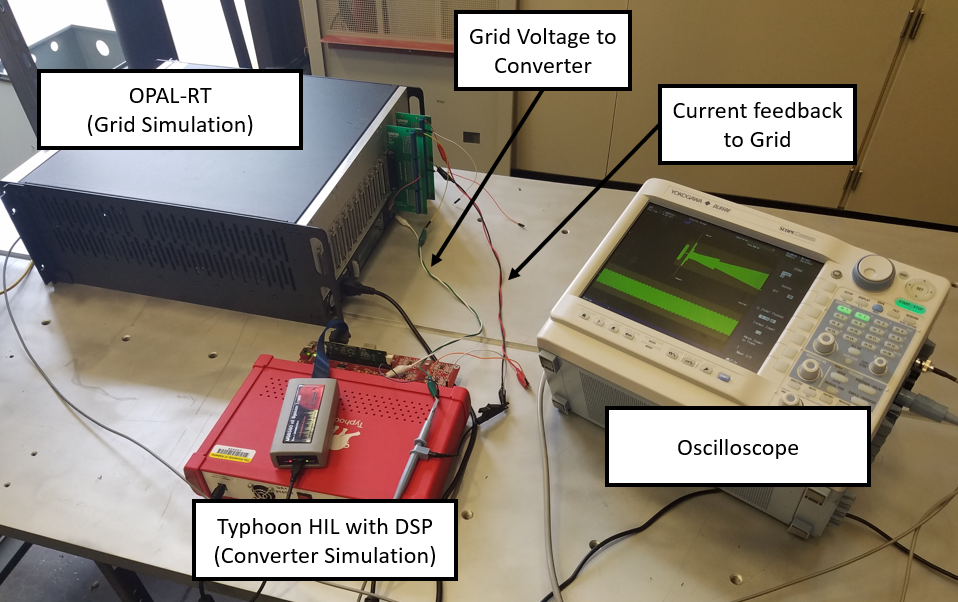}
\vspace{-2mm}
\caption{System hardware set-up.}
\label{fig:SystemTestSetup}
\vspace{-5mm}
\end{figure}

The grid simulator sends a grid voltage signal via the OPAL-RT to the converter and receives instantaneous current feedback from Typhoon HIL device to determine power consumption from that particular node. The actual hardware test system is constructed as seen in Fig.~\ref{fig:SystemTestSetup}.

\begin{figure}[bt]
\centering
\includegraphics[width=85mm]{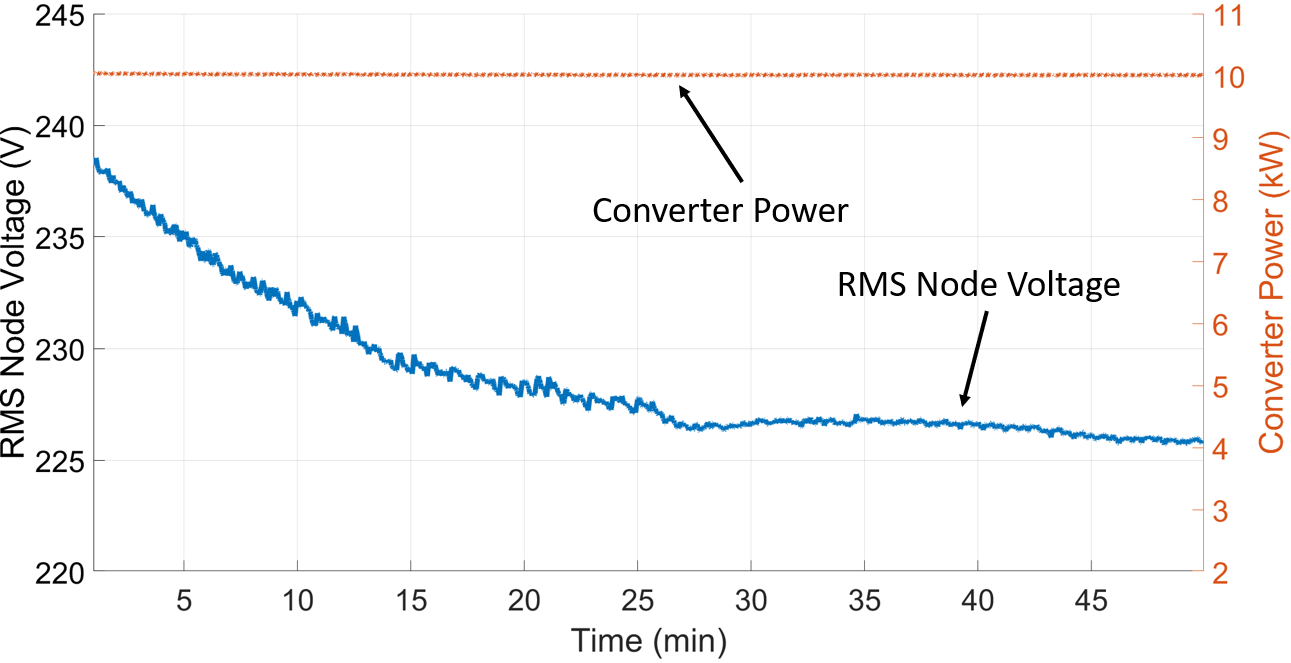}
\vspace{-2mm}
\caption{System test results without AIMD.}
\label{fig:SystemTestWaveformNoAIMD}
\vspace{-6mm}
\end{figure}

This system was tested with the grid simulation starting at approximately 5:30~PM when our load model starts to increase its power demand. The charger operates at a rated power of 10~kW. The system ran for about 50~min, and the results from this test are shown in Fig.~\ref{fig:SystemTestWaveformNoAIMD}. With the increasing load on the grid during the peak hours of the day, the particular end-node voltage has a decreasing trend. The charger does not respond to the increasing loading in the distribution grid and maintains drawing rated power. 

In an attempt to mitigate this voltage drop and reduce the loading on the distribution grid, the additive increase-multiplicative decrease (AIMD) algorithm was implemented into the converter controller~\cite{ucer2019realtime}. Per AIMD, when a decreasing trend in grid voltage is detected, the converter will automatically reduce its power consumption to compensate for the increased feeder congestion. The controller compares the grid voltage to a determined threshold value ($V_{th}$) every 10 seconds (algorithm period, $T_{a}$). This helps avoid very-fast dynamic disturbances of nearby loads that might cause throttling of the converter power if not filtered. If the grid voltage is below the threshold, the converter will multiplicatively decrease its power consumption by scaling it down by $\beta=0.5$; and if the grid voltage is determined to be stable above the threshold, it will additively increase its power consumption by $\alpha=100$~W. Eventually, the charger will converge to an average equilibrium charging power. The algorithm is shown in Alg.~\ref{alg_AIMD}. This algorithm is implemented with a TI DSP. 

\begin{algorithm}[tb]
\caption{Proposed AIMD}\label{alg_AIMD}
\hrule
 \vspace{1mm}
\begin{algorithmic}[1]

 \item[\textbf{Parameter:}] Threshold update period: $T_{u} = 60s$
 \item[\textbf{Parameter:}] Algorithm period: $T_{a} = 10s$
  \item[\textbf{Input:}]Voltage meas. : {$\Vec{V_i}=[V_i(t-T_{u})\cdots V_i(t)]$}

 \item[\textbf{Compute:}] $V_{th} = min(\Vec{V_i})$ at every $t=T_{u}\times k, k\in \mathbb{N} $ 
 \item[\textbf{Parameter:}] Additive increase parameter: $\alpha_{i}=100W$
 \item[\textbf{Parameter:}] Multiplicative decrease parameter: $\beta_{i}=0.5$
 \item[\textbf{Input:}]Previous $P$ consumption: $P_i(t)$
 \item[\textbf{Output:}] New $P$ consumption: $P_i(t+1)$

\item[\textbf{Implement following at every $t=T_{a}\times k, k\in \mathbb{N}$ :}] 
 \IF{$V_i(t) > V_{th}$}
 \STATE $P_i(t+1)=P_i(t) +\alpha_{i}$
 \ELSE
 \STATE $P_i(t+1)=P_i(t) \times \beta_i$
 \ENDIF
\vspace{1mm}
\hrule
\end{algorithmic}
\end{algorithm}


The AC/DC converter uses past grid voltage measurements collected within the previous moving 1-min window (threshold update period, $T_{u}$) to dynamically update the threshold value by assigning it to the minimum recorded voltage in this window. A changing threshold allows the converter to adapt its charging power to the changing loading conditions in the grid. This is critical for a decentralized controller as node voltages vary based on location and other factors.

\begin{figure}[tb]
\vspace{-2mm}
\centering
\includegraphics[width=85mm]{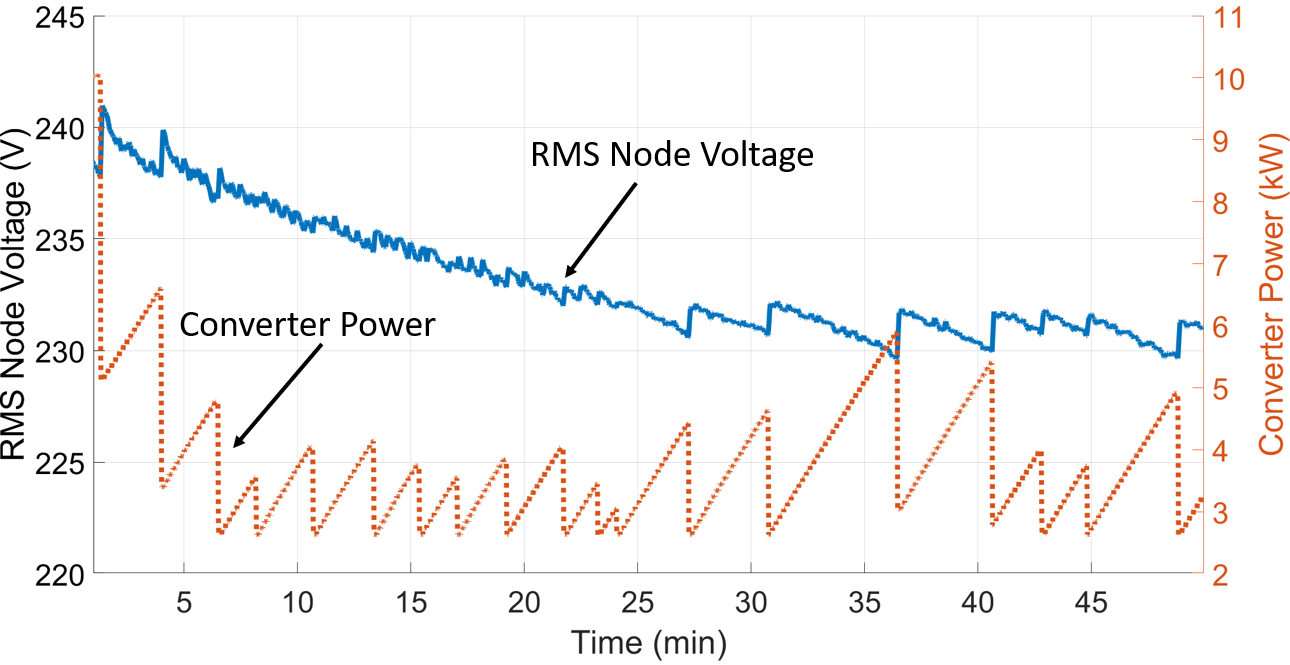}
\vspace{-2mm}
\caption{System test results with AIMD.}
\label{fig:SystemTestWaveformAIMD}
\vspace{-2mm}
\end{figure}

The full system was then retested with the AIMD algorithm implemented, and the results are shown in Fig.~\ref{fig:SystemTestWaveformAIMD} in RMS form. In about 50 minutes of the constant power test, the grid voltage had fallen to approximately 226 $V_{RMS}$ with a charging power of 10~kW. When the AIMD algorithm is used, the grid node voltage only dropped to a minimum of about 229 $V_{RMS}$, but the charger was only able to operate at an average power of 3.9~kW. This 61\% drop in charging power is significant, but was necessary to avoid congestion in the power distribution grid. The algorithm ran without a considerable extra taxation to the TI DSP closed-loop computation, proving its easy implementation at the charger end-node.

\section{Conclusion}
\vspace{-1mm}

In this paper, the design, modeling, and implementation of a controller for a bidirectional AC/DC converter was studied. The HIL environment was used to accelerate the controller development and was tested without the risk of damaging actual hardware. The HIL-based converter was integrated into an external real-time grid simulator to study its impact on the operation of a complex grid model. The entire system was tested with the converter operating at constant rated power and the drop in node voltage was observed. The AIMD algorithm was then implemented in the converter controller as a compromise between node voltage drop and charging power. 

In the future, we plan to construct a hardware version of this converter that will allow for high-power testing, accurate loss measurements, and analysis of the accuracy of the HIL simulations. This converter will also be used in a real time high-power grid testbed, which will allow the full operation of the hardware to be validated and enable accurate experiments of algorithms to optimize grid stability.


\Urlmuskip=0mu plus 1mu
\bibliographystyle{IEEEtran}
\bibliography{./main.bib}

\end{document}